\documentclass[11pt,twoside]{article}

\usepackage{asp2014}

\aspSuppressVolSlug
\resetcounters

\begin{document}

\title{Structuring metadata for the Cherenkov Telescope Array}

\author{Mathieu Servillat$^1$, Catherine Boisson$^1$, Julien Lefaucheur$^1$, Johan Bregeon$^2$, Mich\`ele Sanguillon$^2$ and Jose-Luis Contreras$^3$ for the CTA Consortium$^4$\\
\affil{$^1$Laboratoire Univers et Th\'eories, Observatoire de Paris / PSL Research University / CNRS, Meudon, France; \email{mathieu.servillat@obspm.fr}}
\affil{$^2$Laboratoire Univers et Particules de Montpellier, France}
\affil{$^3$Universidad Complutense de Madrid, Spain}
\affil{$^4$See \url{http://bit.do/cta\_consortium} for full author \& affiliation list}}

\paperauthor{Mathieu Servillat}{mathieu.servillat@obspm.fr}{}{Laboratoire Univers et Th\'eories, Observatoire de Paris / PSL Research University / CNRS}{}{Meudon}{}{92190}{France}
\paperauthor{Catherine Boisson}{}{}{Laboratoire Univers et Th\'eories, Observatoire de Paris / CNRS / PSL}{}{Meudon}{}{92190}{France}
\paperauthor{Julien Lefaucheur}{}{}{Laboratoire Univers et Th\'eories, Observatoire de Paris / CNRS / PSL}{}{Meudon}{}{92190}{France}
\paperauthor{Johan Breg\'eon}{}{}{Laboratoire Univers et Particules de Montpellier}{}{Montpellier}{}{}{France}
\paperauthor{Mich\`ele Sanguillon}{}{}{Laboratoire Univers et Particules de Montpellier}{}{Montpellier}{}{}{France}
\paperauthor{Jose-Luis Contreras}{}{}{Universidad Complutense de Madrid}{}{Madrid}{}{}{Spain}

\begin{abstract}
The landscape of ground-based gamma-ray astronomy is drastically changing with the perspective of the Cherenkov Telescope Array (CTA) composed of more than 100 Cherenkov telescopes. For the first time in this energy domain, CTA will be operated as an observatory open to the astronomy community. In this context, a structured high level data model is being developed to describe a CTA observation. The data model includes different classes of metadata on the project definition, the configuration of the instrument, the ambient conditions, the data acquisition and the data processing. This last part relies on the Provenance Data Model developed within the International Virtual Observatory Alliance (IVOA), for which CTA is one of the main use cases. The CTA data model should also be compatible with the Virtual Observatory (VO) for data diffusion. We have thus developed a web-based data diffusion prototype to test this requirement and ensure the compliance.
\end{abstract}

\section{Cherenkov Astronomy}

The Imaging Atmospheric Cherenkov Technique (IACT) is a method to detect very-high-energy (VHE) gamma-ray photons in the 20 GeV to 300 TeV range. The IACT works by imaging the very short flash of Cherenkov radiation generated by the cascade of relativistic charged particles produced when a very high-energy gamma ray strikes the atmosphere. This is illustrated in Figure~\ref{fig1}.

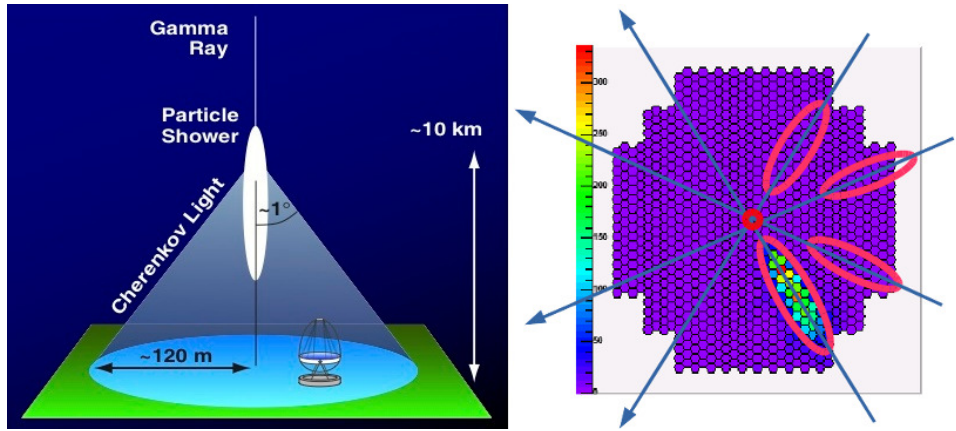
\articlefigure{P5-5_f1.eps}{fig1}{Detection of gamma rays using Cherenkov telescopes. {\it Left}: sketch of a particle shower produced by a gamma ray entering the atmosphere. The Cherenkov light is recorded by one or several telescopes. {\it Right}: image of the shower on the camera of a telescope and basic reconstruction of the event direction for four telescopes.}

The Cherenkov Telescope Array (CTA) project is an initiative to build the next generation ground-based instrument for VHE gamma-ray astronomy. It will provide deep insights into the non-thermal high-energy universe \citep[see e.g.][]{cta_consortium_2013}. Contrary to previous Cherenkov experiments, it will serve as an open observatory to a wide astrophysics community, with the requirement to deliver self-described data products to users that may be unaware of the Cherenkov astronomy specificities.

\section{High Level Data Model}

We show the global structure of the High Level Data Model in Figure~\ref{fig2}, without details on classes and attributes. It is composed of the following parts :

\begin{itemize}

\item \emph{Proposals} are decomposed into \emph{Targets} with their requirements (observing and pointing modes, ...), and constraints (e.g. night sky background, ...);

\item The \emph{Scheduler} then creates an observation program composed of blocks: \emph{Scheduling Blocks} (sequence of observations planned for a given \emph{Target}), made of \emph{Observation Blocks} (effective start and stop times of acquisition with a given configuration);

\item The \emph{Observation Configuration} (Obs Config) defines the coordinates, the {sub-array} (group of telescopes used), the type of observation, the strategy and the observing, pointing and trigger modes;

\item The \emph{Instrument Description} contains the complete instrument description and its modifications;

\item Raw \emph{Data} is produced during \emph{Acquisition} and processed to higher data levels following the different Pipeline stages.

\end{itemize}

\articlefigure{P5-5_f2.eps}{fig2}{Global structure of the High Level Data Model.}

\section{Provenance}

Provenance is information about the generated datasets, the different processing steps performed to obtain them, and people involved in producing them. The tracking of processing stages will be done using the IVOA Provenance Data Model, based on the W3C PROV ontology (Entity-Activity-Agent relations). 
This data model and its access layer are currently in development and a first working draft of the recommendation will be released (see \citealt{riebe_adassxxvi} and \citealt{louys_adassxxvi} in these proceedings for more information, and also \citealt{sanguillon_adassxxv}).

\section{VO Diffusion for CTA}

One of the goals of the High Level Data Model is to make CTA data products available and discoverable through the Virtual Observatory (VO).
For example, the attributes contained in this data model can  be mapped to the generic IVOA ObsCore data model, and exposed using the IVOA Table Access Protocol (TAP). This provides an ObsTAP service for the CTA Archive.

An online prototype has been developped to test the data model and adapt the VO protocols to Cherenkov Astronomy:
\url{https://voparis-cta-test.obspm.fr}.

\acknowledgements ASTERICS (\url{http://www.asterics2020.eu/}) is a project supported by the European Commission Framework Programme Horizon 2020 Research and Innovation action under grant agreement n. 653477; Additional funding was provided by the INSU (Action Sp\'ecifique Observatoire Virtuel, ASOV), the Action F\'ed\'eratrice CTA at the Observatoire de Paris and the Paris Astronomical Data Centre. This paper has gone through internal review by the CTA Consortium.

\bibliography{P5-5}  

\end{document}